\documentclass[a4paper,12pt]{article}
\usepackage[dvips]{graphics}
\usepackage{epsfig}
\usepackage{psfrag}
\usepackage[psamsfonts]{amssymb}
\usepackage{amsmath}
\usepackage{indentfirst}
\usepackage{amssymb}
\usepackage{wrapfig}

\usepackage{cite}

\begin{document}

\title{Comparative statistics of Garman-Klass,  Parkinson, Roger-Satchell and bridge estimators}

\author{S. Lapinova\thanks{National research University ``Higher school of economics'', Russia}, A. Saichev\thanks{ETH Zurich -- Department of Management, Technology and Economics, Switzerland}}
\date{}

\maketitle

\begin{abstract}

Comparative statistical properties of Parkinson, Garman-Klass, Roger-Satchell and bridge oscillation estimators are discussed. Point and interval estimations, related with mentioned estimators are considered
\end{abstract}

\section{Examples of volatility estimators}

Consider dependence on time $t$ of the price $P(t)$ of some financial instrument. As a rule, at discussing of volatility, one consider its logarithm
\[
X(t) := \ln P(t) .
\]
Let point out one of the conventional volatility $V(T)$ definition, which we are using in this paper: It is the variance
\begin{equation}\label{sqvoldef}
V(T):=\bold{Var}\left[Y(t,T)\right] = \bold{E}\left[Y^2(t,T)\right] - \bold{E}^2\left[Y(t,T)\right] .
\end{equation}
of the log-price increment
$Y(t,T) := X(t+T) - X(t)$
within given time interval duration $T$.

Recall, Garman-Klass (G\&K) \cite{Garman1980}, Parkinson (PARK) \cite{PARK1980} and Roger-Satch\-ell (R\&S) \cite{Roger1991} volatility estimators are resting on the high and low values:
\begin{equation}\label{hilidef}
H := \sup_{t'\in(0,T)} Y(t,t') , \qquad L := \inf_{t'\in(0,T)} Y(t,t') .
\end{equation}
Accordingly, PARK estimator is equal to
\begin{equation}\label{Park}
\hat{V}_p :=  \frac{(H-L)^2}{\ln 16} ,
\end{equation}
while G\&K estimator given by expression
\begin{equation}\label{GKest}
\begin{array}{c} \displaystyle
\hat{V}_g := k_1 (H-L)^2-k_2 (C(H-L)- 2 H L) - k_3 C^2 ,
\\
k_1 = 0.511 , \qquad  k_2 = 0.0109 , \qquad k_3 = 0.383 .
\end{array}
\end{equation}
Here $C:= Y(t,T)$ is the close value of the log-price increment.
Recall else R\&S estimator, equal to
\begin{equation}\label{RSest}
\hat{V}_r := H (H-C) +L (L-C).
\end{equation}

Besides of mentioned well-known estimators, we discuss \emph{bridge oscillation estimator}. Below we call it shortly by \emph{bridge estimator}. Before to define it, recall bridge $Z(t,t')$ stochastic process definition. It is equal to
\begin{equation}\label{bridgorgdef}
Z(t,t') := Y(t,t') - \frac{t'}{T} ~ Y(t,T) , \qquad t' \in (0,T) .
\end{equation}
Let introduce high and low of the bridge:
\begin{equation}\label{hilibridgedef}
\mathcal{H} := \max_{t'\in(0,T)} Z(t,t') , \qquad \mathcal{L} := \min_{t'\in(0,T)} Z(t,t') .
\end{equation}
Accordingly, mentioned above bridge volatility estimator given by
\begin{equation}\label{bridsqvoldef}
\hat{V}_b :=  \kappa \left(\mathcal{H}-\mathcal{L}\right)^2 .
\end{equation}
The value of the factor $\kappa$ will be calculated later.

\section{Geometric Brownian motion}

One of conventional models of price stochastic behavior is geometric Brownian motion (see \cite{Jeanblanc2009,Cont2004,{Saichev2010}}). In particular, it is used in theoretical justification of G\&K, PARK and R\&S estimators. Below we discuss statistics of mentioned volatility estimators in frame of geometric Brownian motion model. Namely, we assume that increment of the log-price is of the form
\begin{equation}\label{musigwbrm}
Y(t,T) = \mu T + \sigma B(T) .
\end{equation}
Here $\mu$ is the drift of the price, while $B(t)$ is the standard Brownian motion $B(t)\sim \mathcal{N}(0,t)$.
Factor $\sigma^2$ is the intensity of the Brownian motion.

Recall, Brownian motion posses by self-similar property
\begin{equation}\label{bsimasq}
B(t) \sim \sqrt{T}\, B\left(\frac{t}{T}\right) , \qquad \forall~ T > 0 ,
\end{equation}
where and below sign $\sim$ means identity in law.

\begin{figure}
\begin{center}
\includegraphics[width=0.7\linewidth]{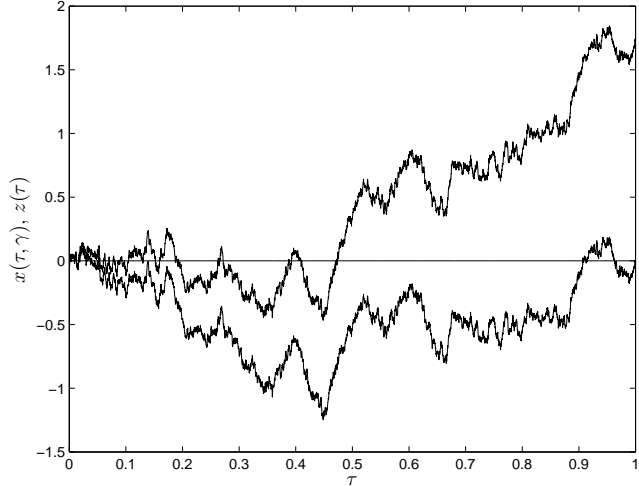}\\
\end{center}
\caption{Typical paths of canonical Brownian motion $x(\tau,\gamma)$ \eqref{xtsimxtau} for $\gamma=1$ and corresponding canonical bridge $z(\tau)$ \eqref{mostzdef}}\label{winbridgam10}
\end{figure}

Using pointed out self-similar property, one can ensure that
\begin{equation}\label{xtsimxtau}
\begin{array}{c}\displaystyle
Y(t,t') \sim \sigma \sqrt{T} ~ x(\tau,\gamma),
\\[4mm] \displaystyle
x(\tau,\gamma):= \gamma \tau + B(\tau) , \qquad
\gamma := \frac{\mu}{\sigma} \sqrt{T} , \qquad
\tau := \frac{t'}{T} \in(0,1) .
\end{array}
\end{equation}
Henceforth we call process $x(\tau,\gamma)$ by \emph{canonical Brownian motion}, while factor $\gamma$ by \emph{canonical drift}.
Using relations \eqref{Park}, \eqref{GKest}, \eqref{bridsqvoldef} and \eqref{xtsimxtau}, one find that
\[
\begin{array}{c}
\hat{V}_p \sim V(T) \cdot \hat{v}_p(\gamma) , \qquad \hat{V}_g \sim V(T) \cdot \hat{v}_g(\gamma) , \qquad \hat{V}_b \sim V(T) \cdot \hat{v}_b ,
\\[2mm]
\hat{V}_r \sim V(T) \cdot \hat{v}_r(\gamma) , \qquad V(T) = \sigma^2 T .
\end{array}
\]
We have used above \emph{canonical estimators}:
\begin{equation}\label{canpgoests}
\begin{array}{c} \displaystyle
\hat{v}_p(\gamma) : = \frac{d^2}{\ln 16} , \qquad \hat{v}_b := \kappa s^2 , \qquad d := h-l , \qquad s := \xi-\zeta ,
\\[4mm]
\hat{v}_g(\gamma) := k_1 d^2-k_2 (c d- 2 h c) - k_3 c^2 , \qquad \hat{v}_r = h (h-c) + l(l-c) ,
\end{array}
\end{equation}
containing high, low and close values
\begin{equation}\label{hlcdef}
h := \sup_{\tau\in(0,1)} x(\tau,\gamma) , \qquad l := \inf_{\tau\in(0,1)} x(\tau,\gamma) , \qquad c := x(1,\gamma) ,
\end{equation}
of canonical Brownian motion, and high and low values
\begin{equation}\label{extrbridge}
\xi := \sup_{\tau\in(0,1)} z(\tau) , \qquad \zeta := \inf_{\tau\in(0,1)} z(\tau) ,
\end{equation}
of the canonical bridge
\begin{equation}\label{mostzdef}
z(\tau) := x(\tau,\gamma)- \tau x(1,\gamma) = B(\tau) - \tau \cdot B(1) , \qquad \tau\in(0,1) .
\end{equation}
Plots of the typical paths of the canonical Brownian motion $x(\tau,\gamma)$ \eqref{xtsimxtau} for $\gamma=1$ and corresponding canonical bridge $z(\tau)$ \eqref{mostzdef} are given in figure~\ref{winbridgam10}.

It is worthwhile to note that the closer expected values of canonical estimators $\hat{v}_p(\gamma)$, $\hat{v}_g(\gamma)$,  $\hat{v}_r$ and $\hat{v}_b$ to unity, the less biased corresponding original volatility estimators. Analogously, the smaller variances of canonical estimators the more efficient original volatility estimators $\hat{V}_p$, $\hat{V}_g$, $\hat{V}_r$ and $\hat{V}_b$.

Notice additionally that canonical drift $\gamma$ of the canonical Brownian motion $x(\tau,\gamma)$ \eqref{xtsimxtau} is, as a rule, unknown. Nevertheless, to get some idea about dependence on drift $\mu$ of bias and efficiency of volatility estimators, we will discuss below in details dependence of canonical estimators statistical properties on possible values of the factor $\gamma$.

\section{Comparative efficiency of PARK and bridge estimators}

Resting on, given at Appendix, analytical formulas for probability density functions (pdfs) of random variables \eqref{hlcdef} and \eqref{extrbridge}, we explore in this section some atatistical properties of canonical PARK estimator $\hat{v}_p(\gamma)$ and bridge one $\hat{v}_b$ \eqref{canpgoests}.

Let check, first of all, unbiasedness of canonical PARK estimator. To make it, let calculate, with help of pdf $q_x(\delta)$ \eqref{qdelexpr}, mean square of oscillation $d=h-l$ of the canonical Brownian motion $x(\tau,\gamma)$ at the zero canonical drift ($\gamma=0$). After simple calculations obtain
\begin{equation}\label{matexphlsq}
\bold{E}[d^2] = 2 + \sum_{m=1}^\infty \frac{2}{m (4 m^2-1)} = \ln 16.
\end{equation}
From here and from expression \eqref{canpgoests} of canonical PARK estimator $\hat{v}_p(\gamma)$ one can see that the following expression is true
\[
\bold{E}[\hat{v}_p(\gamma=0)] = 1 .
\]

Let find now the factor $\kappa$ at expressions \eqref{bridsqvoldef} and \eqref{canpgoests}. To make it, calculate first of all the mean square of the bridge oscillation. Due to expression \eqref{rhodelexp} for the bridge oscillation $s$ \eqref{canpgoests} pdf, one have
\[
\bold{E}[s^2] =  \sum_{m=1}^\infty \frac{1}{m^2} = \frac{\pi^2}{6} .
\]
Accordingly, unbiased canonical bridge estimator has the form
\begin{equation}\label{bgamma6pi}
\bold{E}[\hat{v}_b] = 1 \quad \Rightarrow \quad \kappa =\frac{1}{\bold{E}[s^2]} \quad \Rightarrow \quad
\hat{v}_b = \frac{6\, s^2}{\pi^2} .
\end{equation}

\begin{figure}
\begin{center}
\includegraphics[width=0.7\linewidth]{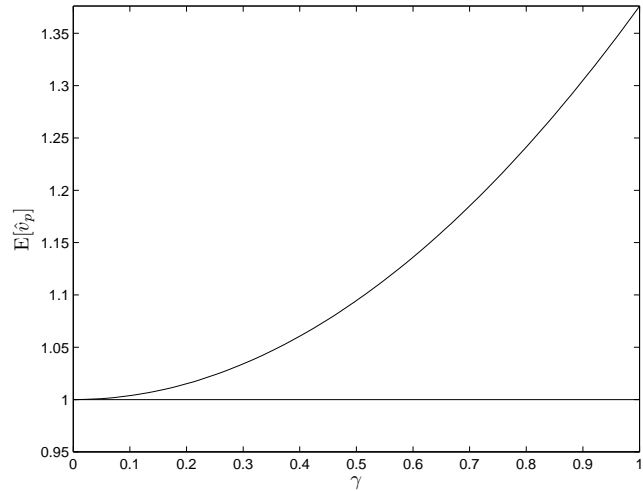}
\end{center}
\caption{Plot of canonical PARK estimator $\hat{v}_p(\gamma)$
mean value, as function of canonical drift $\gamma$. It is seen that with growth of $\gamma$ PARK estimator becomes more and more biased. Straight line is the plot of canonical bridge $\hat{v}_b$, mean value}\label{parkmeangam}
\end{figure}

The great advantage of the bridge estimator is its unbiasedness for any drift. This remarkable property of the pointed out estimator is the consequence of the fact that bridge $Z(t,t')$ \eqref{bridgorgdef} and its canonical counterpart $z(\tau)$ don't depend on the drift $\mu$ (canonical drift $\gamma$) at all. On the contrary, PARK estimator becomes essentially biased at nonzero drift. In figure~\ref{parkmeangam} depicted dependence on $\gamma$ of canonical PARK estimator expected value, illustrating bias of PARK estimator at nonzero drift. Corresponding curve obtained with help of analytical expression \eqref{qdelgamexpr} for canonical bridge oscillation pdf.

Let calculate variances of canonical PARK and bridge estimators. After substitution into the rhs of expression
\[
\bold{E}[\hat{v}^2_p(\gamma=0)] := \frac{1}{\ln^2 16}\int_0^\infty \delta^4 q_x(\delta) d\delta
\]
the sum \eqref{qdelexpr} for the canonical Brownian motion oscillation pdf $q_x(\delta)$, and after summation obtain for $\gamma=0$:
\[
\bold{E}[\hat{v}^2_p(\gamma=0)] = \frac{9 \, \zeta(3)}{\ln^2 16} \simeq 1.40733 .
\]
Accordingly, variance of canonical PARK estimator $\hat{v}_p$ is
\begin{equation}\label{parkvargamzero}
\bold{Var}[\hat{v}_p(0)] = \frac{9 \, \zeta(3)}{\ln^2 16} -1 \simeq 0.407 .
\end{equation}

As the next step, we calculate variance of canonical bridge estimator $\hat{v}_b$ \eqref{bgamma6pi}. Sought variance is equal to
\[
\bold{Var}[\hat{v}_b] := \frac{36}{\pi^4} ~ \bold{E}[s^4] -1 .
\]
After substitution here, following from \eqref{rhodelexp}, relation
\[
\bold{E}[s^4] := \int_0^2 \delta^4 q_b(\delta) d\delta = 3 \sum_{m=1}^\infty \frac{1}{m^4} = \frac{\pi^4}{30} ,
\]
obtain
\begin{equation}\label{varbridge}
\bold{Var}[\hat{v}_b] = \frac{6}{5} -1 = 0.2 .
\end{equation}
Comparing equalities \eqref{parkvargamzero} and \eqref{varbridge},
one can see that variance of bridge estimator approximately twice smaller than variance of PARK estimator.

\begin{figure}
\begin{center}
\includegraphics[width=0.7\linewidth]{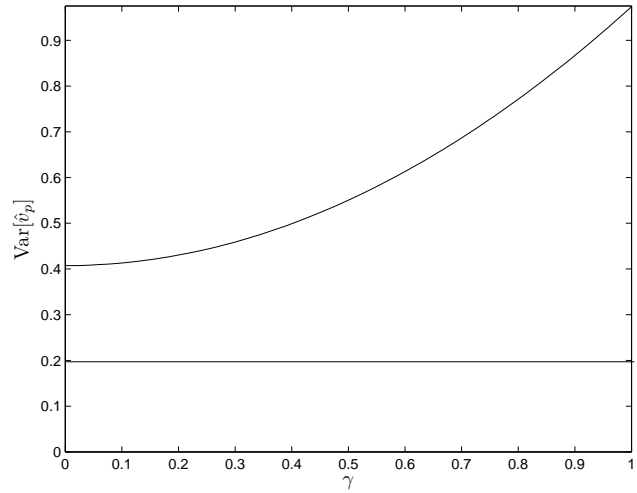}
\end{center}
\caption{Plots of dependence on $\gamma$ of canonical PARK estimator variance. Straight line is the variance of canonical bridge estimator}\label{parkvargam}
\end{figure}

Recall, variance of bridge estimator does not depend on drift. On the contrary, variance of PARK estimator essentially depends on the drift. One can see it in figure~\ref{parkvargam}, where depicted plot of dependence, on canonical drift $\gamma$, of canonical PARK estimator variance.

\begin{figure}
\begin{center}
\includegraphics[width=0.7\linewidth]{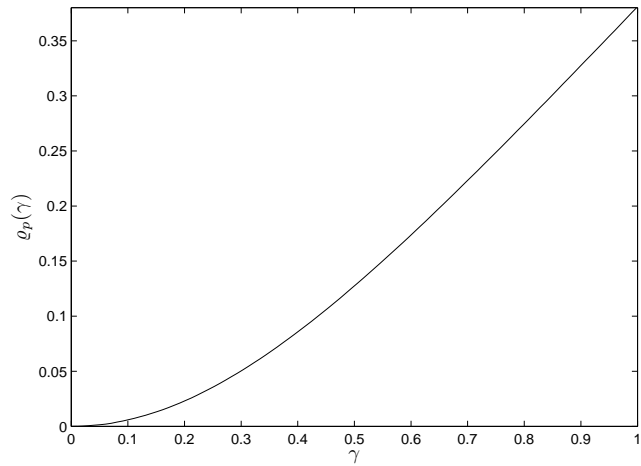}
\end{center}
\caption{Plot of relative bias \eqref{varrhodef} of canonical PARK estimator as function of canonical drift $\gamma$}\label{rhopgam}
\end{figure}

Notice else that bias of some estimator is insignificant only if it is
much smaller than rms of corresponding estimator, i.e. is small the relative bias:
\begin{equation}\label{varrhodef}
\varrho := \frac{\bold{E}[\hat{v}(\gamma)]-1}{\sqrt{\bold{Var}[\hat{v}(\gamma)]}} .
\end{equation}
Plot of canonical PARK estimator relative bias, as function of canonical drift $\gamma$ depicted in figure~~\ref{rhopgam}.

\section{Interval estimations on the basis of PARK and bridge estimators}

\begin{figure}
\begin{center}
\includegraphics[width=0.7\linewidth]{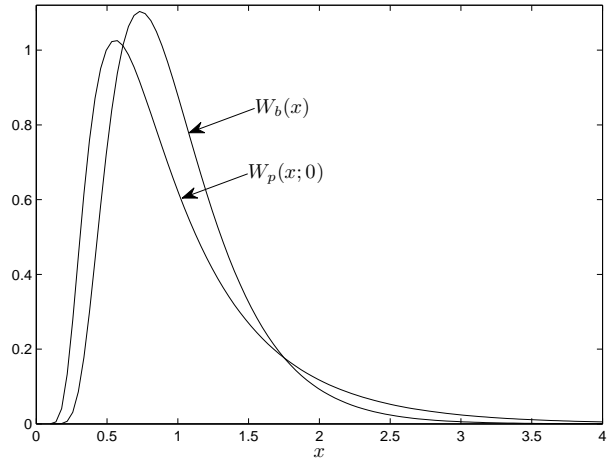}
\end{center}
\caption{Plots of canonical PARK and bridge estimators pdfs, clearly demonstrating ``probabilistic preference'' of bridge estimator in compare with PARK one}\label{pdfspbzero} \end{figure}

\begin{figure}
\begin{center}
\includegraphics[width=0.7\linewidth]{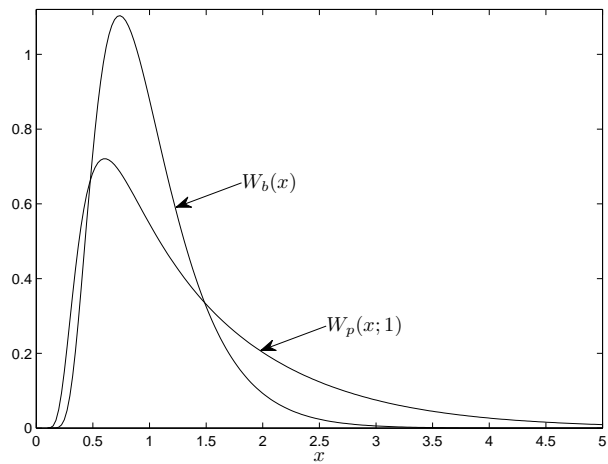}
\end{center}
\caption{Plots of PARK and bridge canonical estimators pdfs for $\gamma=1$}\label{pdfspbone}
\end{figure}

Given at Appendix analytical expressions \eqref{qdelgamexpr}, \eqref{qdelexpr} and \eqref{rhodelexp} for canonical Brownian motion and canonical bridge random oscillations pdfs allow us to explore in details probabilistic properties of PARK and bridge canonical estimators. Let find, at first, pdfs of mentioned canonical estimators random values. It is well known from Probabilistic Theory that pdf $W_p(x;\gamma)$ of canonical PARK estimator is expressed through pdf $q_x(\delta;\gamma)$ \eqref{qdelgamexpr} of canonical Brownian motion oscillation by the relation
\begin{equation}\label{Wpxgam}
W_p(x;\gamma) = \sqrt{\frac{\alpha}{4 x}} ~ q_x\left(\sqrt{\alpha x};\gamma\right) , \qquad \alpha = \ln 16 .
\end{equation}
Similarly, pdf of canonical bridge estimator is equal to
\begin{equation}\label{Wpxbridge}
W_b(x) = \sqrt{\frac{\alpha}{4 x}} ~ q_b\left(\sqrt{\alpha x}\right) , \qquad \alpha = \frac{\pi^2}{6} .
\end{equation}
Here $q_b(\delta)$ \eqref{rhodelexp} is the pdf of canonical bridge oscillation. Plots of canonical PARK estimator pdf, for $\gamma=0$, and pdf of canonical bridge estimator are depicted in figure~\ref{pdfspbzero}. In figure~\ref{pdfspbone} are comparing pdfs of canonical PARK estimator, for $\gamma=1$, and pdf of canonical bridge estimator. It is seen in both figures that pdf of canonical bridge estimator is better concentrated around its expected value $\bold{E}[\hat{v}_b]=1$ than canonical PARK estimator pdf.

Knowing estimators pdfs, one can produce interval estimations of possible volatility values. Consider typical interval estimation:
Let $\hat{V}$ is some volatility estimator, equal to
\begin{equation}\label{hatveqvhatv}
\hat{V} = V(T) \cdot \hat{v} .
\end{equation}
Here $\hat{v}$ is corresponding canonical estimator, while $V(T)$ is the measured volatility. One needs to find probability
\[
F(N) := \bold{Pr}\left\{V(T) < N \cdot \hat{V} \right\}
\]
that unknown (random) volatility $V(T)$ is not more than $N$ times exceeds known (measured) volatility estimated value $\hat{V}$.
It follows from \eqref{hatveqvhatv} that following inequalities are equivalent:
\[
V(T) < N \cdot \hat{V} \qquad \Leftrightarrow \qquad  \hat{v} > 1\big/ N .
\]
Last means in turn that sought probability $F(N)$ is expressed through pdf of canonical estimator  $\hat{v}$ by the following way:
\begin{equation}\label{fnexpr}
F(N) = \bold{Pr}\left\{\hat{v} > 1\big/ N  \right\} = \int_{1/N}^\infty W(x) dx  .
\end{equation}
Here $W(x)$ is the pdf of canonical estimator $\hat{v}$.

Calculations, resting on relations \eqref{Wpxgam}, \eqref{Wpxbridge}, \eqref{fnexpr} give probability $F_b(2)\simeq 0.918$ that true volatility is less than twice of given bridge volatility estimator value $\hat{V}_b$. It is substantially larger than analogous probability in the case of PARK estimator:
$F_p(2,\gamma=0)\simeq 0.813$.

\begin{figure}
\begin{center}
\includegraphics[width=0.7\linewidth]{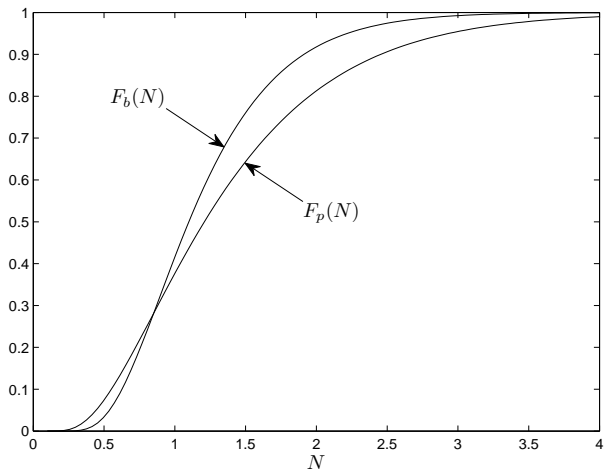}
\end{center}
\caption{Plots of probabilities $F_p(N)$ and $F_b(N)$ that true volatility is less than $N$ times exceeds values of PARK and bridge estimators}\label{fleveln}
\end{figure}

Plots of probabilities $F(N)$ \eqref{fnexpr} dependence on the level $N$, for PARK estimator (in the case of zero drift $\mu=0$) and for bridge volatility estimator are given in figure~\ref{fleveln}.

\section{Comparative statistics of canonical estimators}

Above, we explored in detail statistical properties of two, PARK and bridge estimators. Here we compare their statistics and statistics of another well-known volatility estimators: G\&K and R\&S one. Des\-pite to previous chapters, where we have used known analytical expressions for pdfs of canonical PARK and the bridge estimators, below we use predominantly results of numerical simulations.

Namely, we produce $M\gg1$ numerical simulations of random sequences
\begin{equation}\label{xngamdisc}
x_n(\gamma) := \gamma \frac{n}{N} + \frac{1}{\sqrt{N}} \sum_{n=1}^N \epsilon_n , \qquad n= 0, 1,\dots, N, \qquad x_0(\gamma) = 0 ,
\end{equation}
where $\{\epsilon_n\}$ are iid Gaussian variables $\sim\mathcal{N}(0,1)$. Notice that stochastic process $x_n(\gamma)$ of discrete argument $n$ rather accurately approximates, for large $N\gg1$, paths of canonical Brownian motion $x(\tau,\gamma)$ \eqref{xtsimxtau}.

\begin{figure}
\begin{center}
\includegraphics[width=0.9\linewidth]{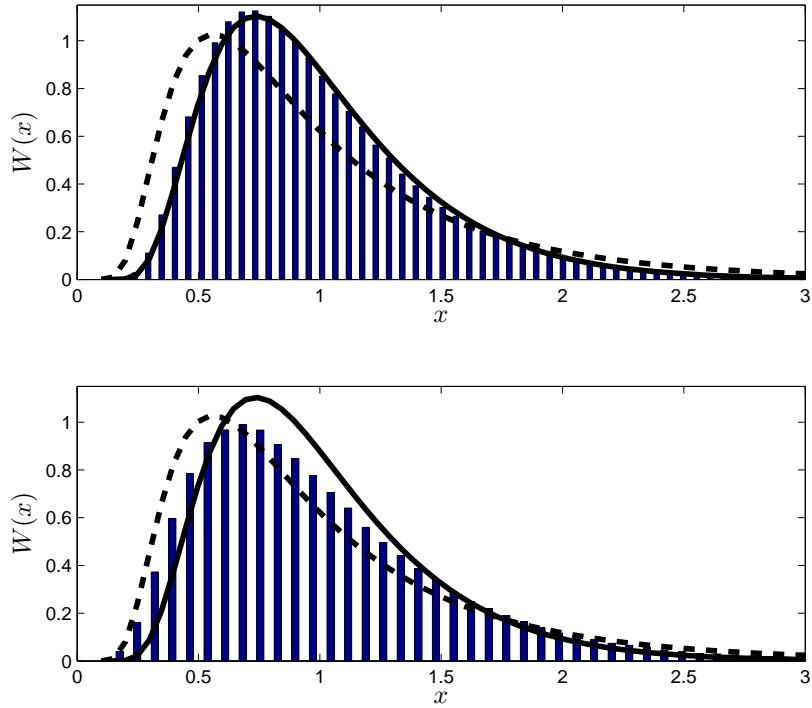}
\end{center}
\caption{\textbf{Upper panel:} Histogram of $M$ samples of canonical bridge estimator $\hat{v}_b$. Solid line is the plot of canonical bridge estimator's pdf, given by analytical expression \eqref{Wpxbridge}, \eqref{rhodelexp}. Dashed line is the pdf of canonical PARK estimator for $\gamma=0$.
\textbf{Lower panel:} Histogram of $M$ samples of canonical G\&K estimator $\hat{v}_g$ for $\gamma=0$. Solid line is the plot of the canonical bridge estimator pdf. Dashed line is the canonical PARK estimator pdf for $\gamma=0$}\label{bargkpdfs}
\end{figure}

\begin{figure}
\includegraphics[width=0.99\linewidth]{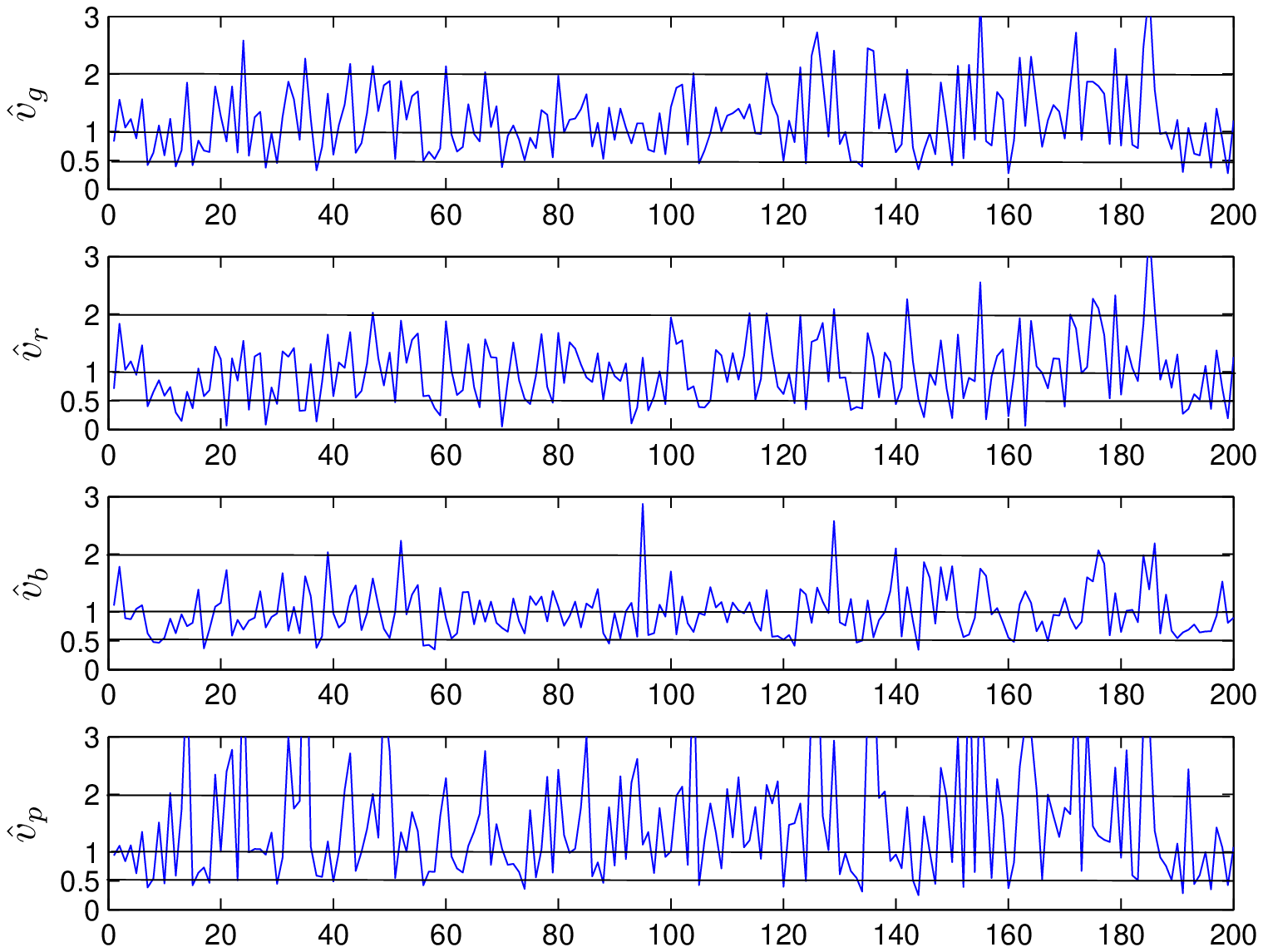}\\
\caption{Plots of two hundreds samples of canonical estimators. Up to down are samples of G\&K, R\&S, bridge and PARK estimators. It is seen even by ``naked eye'' that bridge estimator estimates volatility more accurately than another mentioned estimators}\label{hotvsamples}
\end{figure}

Knowing $M$ iid sequences $\{x_n(\gamma)\}$ one can find corresponding iid samples of pointed out above canonical estimators. Everywhere below we take number of iid samples $M$ and discretization number $N$ equal to
\[
N = 5\cdot 10^3 , \qquad M = 5 \cdot 10^5 .
\]
Plots in figure~\ref{bargkpdfs} demonstrate rather convincingly accuracy of numerical simulations. In figure~\ref{hotvsamples}
are given two hundred samples of canonical G\&K and bridge estimators, ensuring ``by naked eye'' that canonical bridge estimator is more efficient than G\&K one.

\begin{figure}
\begin{center}
\includegraphics[width=0.75\linewidth]{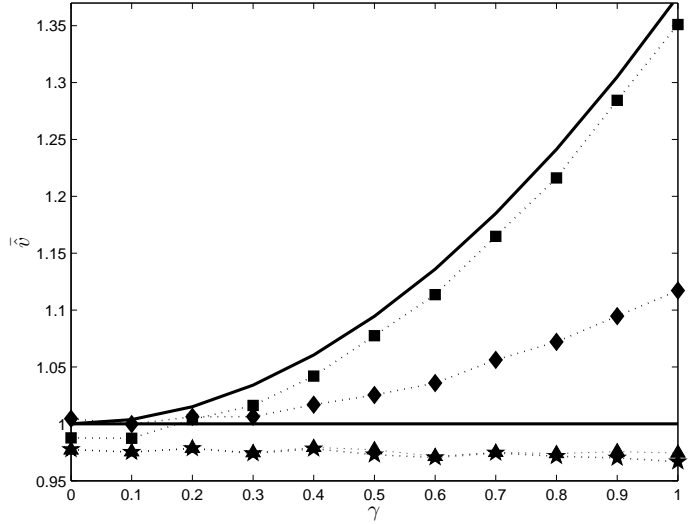}
\end{center}
\caption{Mean values $\bar{\hat{v}}$ of canonical PARK  ($\blacksquare$), G\&K ($\blacklozenge$), R\&S ($\bigstar$) and bridge ($\blacktriangle$) estimators.
Solid lines are theoretical expectations, borrowing from figure~\ref{parkmeangam}}\label{estmeandats}
\end{figure}

In figure~\ref{estmeandats} are given, obtained by numerical simulations, plots of canonical G\&K, PARK, R\&S and bridge estimators mean values, illustrating bias of G\&K and PARK estimators for nonzero canonical drift $\gamma\neq 0$, and actual absence of bias for bridge and R\&S estimators.

\begin{figure}
\includegraphics[width=0.98\linewidth]{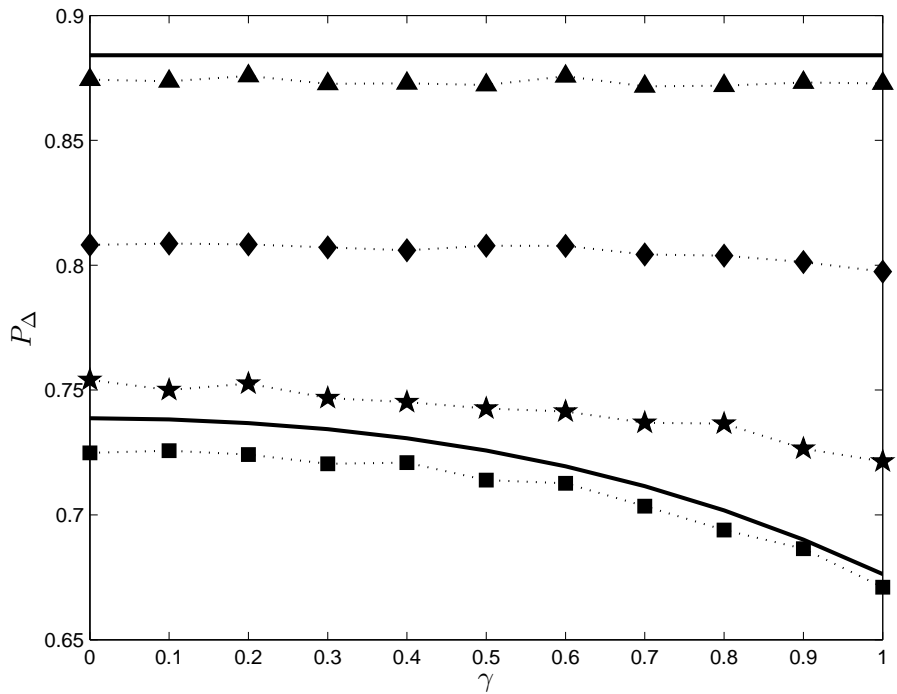}\\
\caption{Estimations of probability $P_\Delta$ \eqref{pdeldef} at different $\gamma$ values, for PARK
($\blacksquare$), R\&S ($\bigstar$), G\&K ($\blacklozenge$) and bridge ($\blacktriangle$) estimators. Solid lines are results of theoretical calculations, resting on formula \eqref{pdeldef}}\label{pdel}
\end{figure}

Eventually, in figure~\ref{pdel} are given plots of probabilities that true volatility $V(T)$ is larger than half of corresponding estimator value and less than twice of it:
\begin{equation}\label{pdeldef}
P_\Delta := \bold{Pr}\left\{\frac{\hat{V}}{2} <  V(T)< 2 \hat{V}\right\} = \int_{1/2}^2 W(x) dx   .
\end{equation}
It is seen that for any $\gamma$ mentioned probability is essentially larger for bridge estimator, than for G\&K, R\&S and PARK estimators.

\section{Acknowledgements}

We are grateful for scientific and financial help of Higher school of economics (Russia, Nizhny Novgorod) and Nizhny Novgorod State University (Russia).

\clearpage

\begin{figure}
\begin{center}
\includegraphics[width=0.7\linewidth]{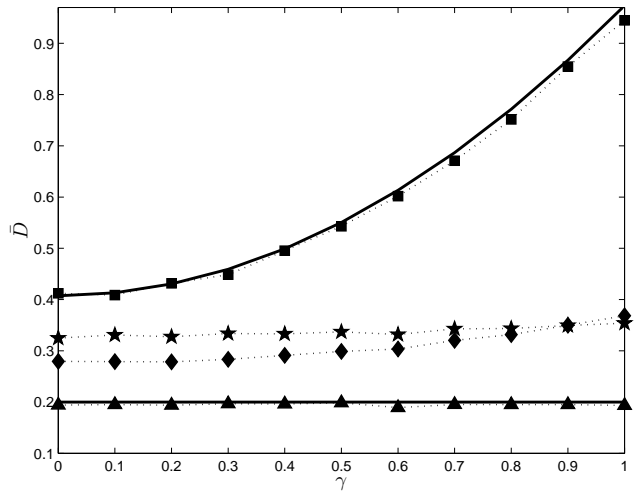}
\end{center}
\caption{Estimations $\bar{D}$ of variance of PARK ($\blacksquare$), R\&S ($\bigstar$), G\&K ($\blacklozenge$) and bridge ($\blacktriangle$) canonical estimators. Solid lines are plots of theoretical variances, borroved from the figure~\ref{parkvargam}. It is seen that for any $\gamma$ bridge estimator's variance significantly smaller than variances of another mentioned estimators}\label{estvardats}
\end{figure}

\clearpage

\appendix
\setcounter{section}{0}
\setcounter{equation}{0}
\renewcommand{\theequation}{\thesection.{\arabic{equation}}}
\renewcommand{\thesection}{\Alph{section}}

\section{Probabilistic properties of high, low and close values}

Here are given pdfs of random variables $(h,l,c)$ \eqref{hlcdef} and variables $(\xi,\zeta)$ \eqref{extrbridge}, which one need for canonical estimators \eqref{canpgoests} statistical analysis.
Let begin with random variable $c=x(1,\gamma)$. Obviously, its pdf is
\[
f(\chi;\gamma) := \frac{1}{\sqrt{2\pi}} \exp\left(-\frac{(\chi-\gamma)^2}{2} \right) , \qquad \chi\in(-\infty,\infty) .
\]
It is easy to show, additionally, that joint pdf  $q_x(\eta,\chi;\gamma)$ of high value $h$ \eqref{hlcdef} of canonical Brownian motion $x(\tau,\gamma)$ and the close value $c=x(1,\gamma)$ is equal to
\begin{equation}\label{etachi}
\begin{array}{c} \displaystyle
q_x(\eta,\chi;\gamma) = \sqrt{\frac{2}{\pi}} \, (2\eta-\chi) \, e^{2\gamma \eta} \exp\left( - \frac{1}{2} (2\eta - x + \gamma)^2 \right) ,
\\[4mm] \displaystyle
\chi < \eta , \qquad \eta >0.
\end{array}
\end{equation}
In turn, pdf of high value $h$ \eqref{hlcdef}
\[
q_x(\eta;\gamma) := \int_{-\infty}^h q_x(\eta, \chi;\gamma) d\chi
\]
given by expression
\begin{equation}\label{qhagmdist}
q_x(\eta;\gamma) = \sqrt{\frac{2}{\pi}} \exp\left(- \frac{(\eta-\gamma)^2}{2}\right) - \gamma e^{2\gamma \eta} \, \text{erfc}\left(\frac{\eta+\gamma}{2}\right) , \qquad \eta > 0 .
\end{equation}

Let write here explicit expression for joint pdf $q_x(\eta,\ell,\chi;\gamma)$ of random variables $(h,l,c)$ \eqref{hlcdef}.
Using formulas, given at the monograph \cite{Borodin2002} and in the article \cite{SaichevArx}, one might show that pointed out joint pdf given by:
\begin{equation}\label{hljpdfexpr}
\begin{array}{c} \displaystyle
q_x(\eta,\ell,\chi;\gamma) = f(\chi;\gamma) \, \mathcal{S}(\eta,\ell|\chi) ,
\\[3mm] \displaystyle
\chi \in (\ell, \eta), \qquad h> \chi \bold{1}(\chi) , \qquad \ell < \chi \bold{1}(-\chi) .
\end{array}
\end{equation}
Here $\bold{1}(\chi)$ is the unit step function, equal to unity for $\chi>0$ and zero otherwise. Besides, above there is function
\begin{equation}\label{mathsdef}
\begin{array}{c}
\mathcal{S}(\eta,\ell|\chi) :=
\\[1mm] \displaystyle
\sum_{m=-\infty}^\infty m \left[ m \mathcal{F}(m(\eta-\ell),\chi)+(1-m) \mathcal{F}(m (\eta-\ell)+\ell,\chi) \right] ,
\\[3mm] \displaystyle
\mathcal{F}(\eta,\chi) := \left[(\chi-2 \eta)^2-1 \right] e^{2 \eta (\chi-\eta)} .
\end{array}
\end{equation}

We need, at exploring statistical properties of canonical G\&K estimator, in joint pdf $q_x(\delta,\chi;\gamma)$ of canonical Brownian motion $x(\tau,\gamma)$ \eqref{xtsimxtau} oscillation $d=h-l$ and the close value $c=x(1,\gamma)$. As it follows from \eqref{hljpdfexpr}, \eqref{mathsdef}, mentioned pdf is equal to
\begin{equation}\label{qetchiexpr}
\begin{array}{c} \displaystyle
q_x(\delta,\chi;\gamma)= 4 f(\chi;\gamma)\sum_{m=-\infty}^\infty m \times
\\[4mm] \displaystyle
\left[ m (\delta-|\chi|) [(|\chi|+2m\delta)^2-1]-(m+1) (|\chi|+2m\delta)\right] e^{-2m\delta(|\chi|+m\delta)} ,
\\[4mm] \displaystyle
\delta> |\chi| , \qquad \chi \in (-\delta,\delta) .
\end{array}
\end{equation}

After integration above joint pdf over all $\chi$ values obtain pdf $q_x(\delta;\gamma)$ of oscillation $d$:
\begin{equation}\label{qdelgamexpr}
\begin{array}{c} \displaystyle
q_x(\delta;\gamma) = \sum_{m=-\infty}^\infty m \bigg[ \sqrt{\frac{8}{\pi}} \exp\left(-\frac{(1+2 m)^2 \delta^2 + 2 \delta\gamma+\gamma^2}{2} \right) \times
\\[4mm] \displaystyle
\bigg(2 \exp\left(\frac{\delta (\delta+4 m \delta+2\gamma)}{2}\right) (2 m^2 \delta^2-1-m (2+\gamma^2)) +
\\[3mm] \displaystyle
(1+e^{2\delta \gamma}) (1+m (2+\gamma^2))\bigg) -2 \gamma \big(a(\delta,\gamma,m) -a(\delta,-\gamma,m)\big) \bigg] ,
\\[2mm]
\delta> 0 .
\end{array}
\end{equation}
Here have used auxiliary function
\[
\begin{array}{c} \displaystyle
a(\delta,\gamma,m) := e^{2 m \delta \gamma}\left[1+m(3+\gamma (\delta + 2m \delta+\gamma))\right] \times
\\[4mm] \displaystyle
\left[ \text{erf}\left(\frac{2 m \delta + \gamma}{\sqrt{2}}\right) - \text{erf}\left(\frac{\delta+2 m\delta +\gamma}{\sqrt{2}} \right) \right], \qquad \delta> 0 .
\end{array}
\]
In particular case of zero drift ($\gamma=0$), one get from \eqref{qdelgamexpr} following expression
\begin{equation}\label{qdelexpr}
\begin{array}{c} \displaystyle
q_x(\delta) =
\\[3mm]\displaystyle
\sqrt{\frac{8}{\pi}} \sum_{m=-\infty}^\infty \left[ (2 m-1)^2 \exp\left(-\frac{(2 m-1)^2 \delta^2}{2}\right) - 4 m^2 e^{-2 m^2 \delta^2} \right] ,
\\[4mm]
\delta> 0 .
\end{array}
\end{equation}

All statistical properties of high and low values \eqref{extrbridge} of canonical bridge \eqref{mostzdef} are defined by their two-fold joint pdf $q_b(\eta,\ell)$, given by relation
\begin{equation}\label{helbridgepdf}
\begin{array}{c} \displaystyle
q_b(\eta,\ell) = \sum_{m=-\infty}^\infty m \left[ m \mathcal{F}(m (\eta - \ell))+ (1-m) \mathcal{F}(m(\eta-\ell)+\ell) \right] ,
\\[4mm] \displaystyle
\mathcal{F}(\eta) := 4 (4 \eta^2 -1) e^{-2 \eta^2} .
\end{array}
\end{equation}
Following from here pdf  $q_b(\delta)$ of canonical bridge oscillation $s= \xi-\zeta$ given by equality
\begin{equation}\label{rhodelexp}
q_b(\delta) = 8 \delta \sum_{m=1}^\infty m^2 (4 m^2 \delta^2 - 3) e^{-2 m^2 \delta^2} , \qquad \delta>0 .
\end{equation}

\clearpage

\end{document}